\documentclass[11pt,a4paper]{article}
\pdfoutput=1
\usepackage{jinstpub}
\usepackage{siunitx,algpseudocode,algorithm,array,booktabs,bm,multirow,makecell,placeins,etoolbox}
\graphicspath{{figures/}}
\newcommand{\ISE}{I_{\mathrm{SE}}}
\newcommand{\ILE}{I_{\mathrm{LE}}}
\newcommand{\predLE}{\hat{I}_{\mathrm{LE}}}
\AtBeginEnvironment{algorithmic}{\small}
\title{Residual diffusion for fine-structure enhancement in low-count accelerator-based muon imaging}
\author[a]{Haochen Wang,}
\author[b,c]{Pei Yu,}
\author[b,c,1]{Zhengyang He\note{Corresponding authors.},}
\author[b,c,d,e]{Yuhong Yu,}
\author[b,c,d,e]{Xueheng Zhang,}
\author[b,c,d,e,1]{Liangwen Chen,}
\author[a,1]{Yu Zhang,}
\author[b,c,d,e]{Lei Yang,}
\author[b,c,d,e,1]{and Zhiyu Sun}
\affiliation[a]{School of Physics, Hefei University of Technology, Hefei 230601, China}
\affiliation[b]{Advanced Energy Science and Technology Guangdong Laboratory, Huizhou 516000, China}
\affiliation[c]{Institute of Modern Physics, Chinese Academy of Sciences, Lanzhou 730000, China}
\affiliation[d]{State Key Laboratory of Heavy Ion Science and Technology,\\Institute of Modern Physics, Chinese Academy of Sciences, Lanzhou 730000, China}
\affiliation[e]{School of Nuclear Science and Technology, University of Chinese Academy of Sciences, Beijing 100049, China}
\emailAdd{hzy@gdlhz.ac.cn}
\emailAdd{chenlw@impcas.ac.cn}
\emailAdd{dayu@hfut.edu.cn}
\emailAdd{sunzhy@impcas.ac.cn}
\abstract{Accelerator-based muon imaging combines the penetrating power of high-energy muons with selectable momentum and a controlled beam direction, offering a powerful means of probing internal structures in large, dense and heavily shielded objects. Within a fixed beam-time budget, realizing this potential requires reliable recovery of fine structures from limited muon statistics. We propose an observation-conditioned residual diffusion method with decoupled linear schedules to enhance low-count muon images through iterative restoration guided by the observed image. We evaluate a simplified VSC-24 fuel cask under an idealized 5 GeV parallel muon beam, using 500,000 accepted events per low-count image and 40 million per reference image. Simulation results show improved recovery of structural boundaries and fine details, with the mean peak signal-to-noise ratio increasing from 11.4239 to 22.1719 dB. These findings highlight the potential of diffusion models to make more effective use of limited muon data and broaden the application of accelerator-based muon imaging across scientific and industrial settings.}
\hypersetup{pdftitle={Residual diffusion for fine-structure enhancement in low-count accelerator-based muon imaging},pdfauthor={Haochen Wang, Pei Yu, Zhengyang He, Yuhong Yu, Xueheng Zhang, Liangwen Chen, Yu Zhang, Lei Yang, Zhiyu Sun}}
\begin{document}
\maketitle
\flushbottom
\section{Introduction}
Muon scattering tomography (MST) reconstructs internal material distributions from the deflection of tracked muons \cite{Borozdin2003}. Accelerator-based MST combines the penetration of GeV muons with selectable momentum and a controlled beam direction, enabling detailed inspection of large, dense objects behind substantial shielding \cite{Suerfu2016}. A feasibility study of the proposed muon source at the High Intensity Heavy-Ion Accelerator Facility (HIAF) reports a maximum muon transport momentum of 7.5 GeV/$c$ and predicts intensities of order $10^5$ muons/s in a purification scenario at 1.5 GeV/$c$ \cite{Xu2025HIAF}.

Realizing this imaging potential requires enough detected events to resolve local scattering differences within the available beam time for each object. At a limited event count, point-of-closest-approach (PoCA) images contain strong statistical fluctuations that obscure material boundaries and narrow gaps. Longer acquisitions improve boundary and gap visibility by accumulating more scattering information. Within a fixed beam-time budget, image enhancement aims to recover structural detail from the available events, with the prospect of reducing the acquisition time and experimental resources required for detailed imaging.

Muon computed-tomography simulations of a VSC-24 cask have examined missing fuel assemblies and substitution by other materials \cite{Liu2017VSC24}. More recently, the Bayesian trajectory-based $\mu$TRec method improved fuel-cask reconstructions over PoCA in simulations with limited muon counts \cite{Ughade2025MuTRec}.

Using paired low- and high-count images, deep learning models can learn to map statistically noisy reconstructions to high-count references \cite{ODonnell2025}. U-Net methods have been used for muon image enhancement in simulated structural diagnostics and experimental PoCA studies \cite{Pezzotti2025,Wang2026}. For fine-structure enhancement, this mapping must recover both local boundaries and the spatially varying texture of high-count scattering images. Diffusion models offer an iterative formulation of this restoration task through learned denoising \cite{SohlDickstein2015,Ho2020}. Within this formulation, residual-shifting and residual-denoising approaches incorporate the difference between degraded and target images into the restoration process \cite{yue2023,liu2024}. Building on these approaches, we centre the terminal distribution on the observed low-count image and predict the residual between this image and a high-count reference. Separate linear schedules control residual attenuation and noise growth, allowing structural correction and stochastic perturbation to evolve gradually along the restoration trajectory.

We demonstrate the method using a simplified VSC-24 fuel cask. Imaging the cask requires both penetration through a metre-scale object with thick shielding and resolution of boundaries within a repeated fuel-block arrangement. These features make it a representative complex object for evaluating fine-structure recovery. Under an idealized 5 GeV parallel muon beam, we enhance short-exposure (SE) PoCA images reconstructed from $5\times10^5$ accepted events and evaluate the outputs against long-exposure (LE) references reconstructed from $4\times10^7$ accepted events. Controlled comparisons examine the effects of schedule decoupling, noise perturbation and iterative restoration. We then compare U-Net and Restormer within the same framework and evaluate recovery in four mixed removal and replacement scenes. Together, these comparisons establish the method's ability to improve image quality at a specified event budget and its value for efficient accelerator-based muon imaging.

\section{Simulation setup and evaluation protocol}
\label{sec:data}
\subsection{Simulation and PoCA reconstruction}

\begin{figure}[tbp]
\centering
\includegraphics[width=130mm]{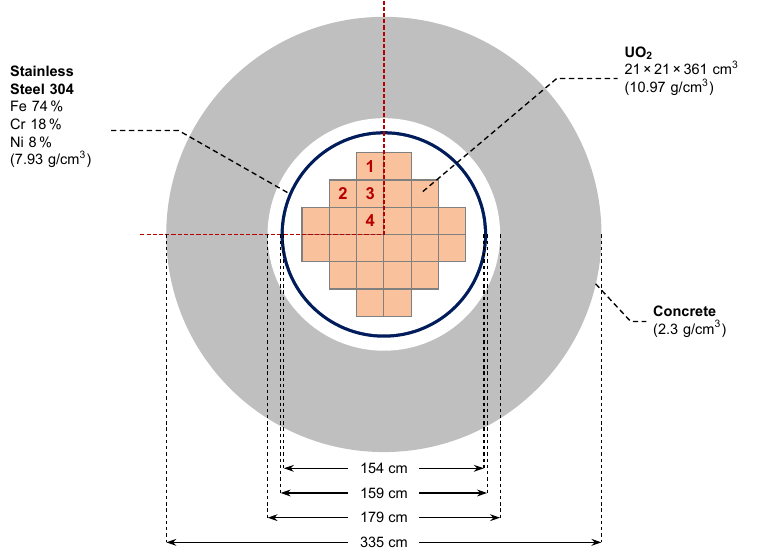}
\caption{Bottom cross-section of the modelled VSC-24 cask. The model is 361 cm tall, with 2.5 cm thick stainless steel 304 and 41.5 cm thick concrete covers at the top and bottom. Dimensions and material labels define the simplified fuel-block geometry used in this study.}
\label{figure1}
\end{figure}

A simplified VSC-24 spent-fuel storage cask was selected to evaluate the recovery of internal boundaries in a large, strongly scattering object \cite{Anigstein2004}. Its geometry and material composition are shown in figure~\ref{figure1}. The model contains 24 fuel blocks within steel and concrete shielding; each block is represented by a $21\times21\times361$ cm$^3$ volume of UO$_2$. Muon transport and detector responses were simulated using Geant4 \cite{Agostinelli2003}. Two position-sensitive detector planes were placed above the cask and two below it to reconstruct the incoming and outgoing muon tracks. A \SI{5}{GeV} monoenergetic parallel muon beam was used as an idealized controlled-source benchmark. Images were reconstructed from the detected events using the PoCA algorithm, following the procedure described in our previous work \cite{Wang2026}. A central region of $\SI{170}{cm}\times\SI{170}{cm}$ was selected as the field of view for reconstruction. The voxel edge length was set to $\SI{2.5}{mm}$, yielding reconstructed images with $680\times680$ pixels. Only events that were detected by all four tracking planes and had a reconstructed scattering angle below $35^\circ$ were retained for PoCA reconstruction. The scattering angle is defined as the three-dimensional angle between the incoming and outgoing track directions. The upper-angle cut was used to suppress noise spots with high PoCA values produced by extreme-angle events. This selection was applied before forming the SE and LE images, and the event budgets reported below refer to the accepted events.

\subsection{Scene configurations and image pairs}
The dataset includes both removal (RM) and replacement (RP) anomalies for individual modelled fuel blocks and entire rows. Iron (Fe) was used as the replacement material. Owing to the symmetric geometry of the VSC-24 cask, single-block anomalies were simulated at the four representative positions shown in figure~\ref{figure1}; images for the remaining positions were generated by rotation and flipping. The same augmentation strategy was applied to row-wise anomalies, for which only the first three representative rows were simulated. In total, 72 anomaly scenes were constructed, comprising 24 single-block RM, 24 single-block RP, 12 row-wise RM, and 12 row-wise RP scenes. Of these scenes, 64 were used for training and eight for validation.

Each scene has an LE reference reconstructed from 40,000,000 accepted muon events and five SE images reconstructed from 500,000 accepted events each, using different random seeds. The five SE images share the same LE reference. The five SE realizations expose the model to different statistical fluctuations for the same underlying scene. This yielded $72\times5$ LE-SE pairs for training and validation. Each original simulated scene, its five SE realizations and paired LE references, and all images derived from them by rotation or flipping were assigned together to either the training or validation set. The SE and LE labels correspond to the low- and high-count image sets defined by these accepted-event budgets. Evaluation uses a complete scene without anomalies and four mixed scenes, MS1--MS4, containing combinations of RM and RP anomalies. The complete scene is used for controlled comparisons of the restoration components and network backbones; the four mixed scenes assess generalization to mixed removal and replacement configurations (section~\ref{ssec:mx}).

\begin{figure}[tbp]
\centering
\includegraphics[width=98mm]{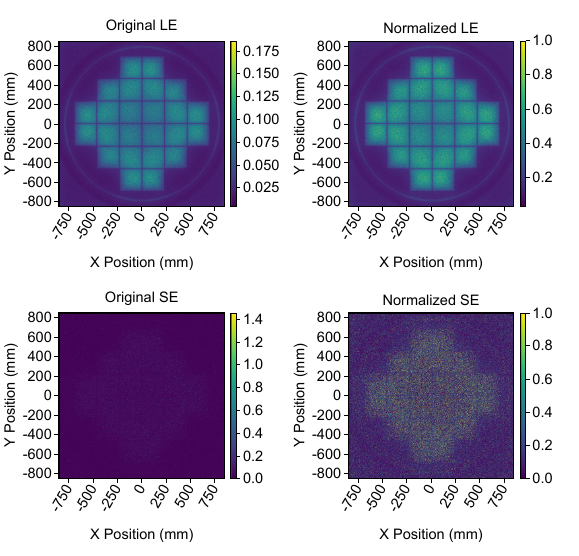}
\caption{Comparison of the SE and LE PoCA-reconstructed images before and after normalization.}
\label{figure2}
\end{figure}

\subsection{Image normalization}
The paired images must share an intensity scale for their residual to represent an intensity correction on the same scale. We therefore normalized all LE and SE images using a common global scale (figure~\ref{figure2}). The normalization factor was defined as
\begin{equation}
V_{\max}^{\mathrm{LE}} = 1.1 \times \max_{i=1,\dots,50} \Bigl\{ P_{99.9}\bigl(I_{\mathrm{LE},i}^{\mathrm{raw}}\bigr) \Bigr\},
\end{equation}
where $P_{99.9}$ denotes the intensity threshold below which $\SI{99.9}{\%}$ of pixel values fall. Here, $I_{\mathrm{LE},i}^{\mathrm{raw}}$ denotes the $i$th unnormalized LE image. This percentile-based scale reduces the influence of isolated high-intensity pixels. The maximum percentile value among 50 randomly selected LE images from the training set was multiplied by 1.1 to provide an additional margin. All LE and SE images were then normalized as
\begin{equation}
\label{g}
x_\mathrm{norm} =
\begin{cases}
\dfrac{x}{V_{\max}^{\mathrm{LE}}}, & 0 \le x \le V_{\max}^{\mathrm{LE}} \\
1, & x > V_{\max}^{\mathrm{LE}}.
\end{cases}
\end{equation}
Here, $x$ and $x_{\mathrm{norm}}$ denote the pixel values before and after normalization, respectively. The same $V_{\max}^{\mathrm{LE}}$ was applied to both LE and SE images, placing them on a common intensity scale defined by the more statistically stable LE images. For pixels below the clipping threshold, this scaling preserves relative intensity differences. With clipping included, the normalized residual is
\begin{equation}
r_{\mathrm{norm}} = g(\ILE^{\mathrm{raw}})-g(\ISE^{\mathrm{raw}}),
\end{equation}
where $g(x)$ denotes the normalization and clipping function in eq.~\eqref{g}, and the superscript raw denotes unnormalized pixel values. For unclipped input pixels, this residual equals $(\ILE^{\mathrm{raw}}-\ISE^{\mathrm{raw}})/V_{\max}^{\mathrm{LE}}$. Hereafter, $\ILE$ and $\ISE$ denote the normalized images and $r=\ILE-\ISE$.

\subsection{Evaluation protocol}
The evaluation comprises controlled comparisons of the restoration components, a comparison of network backbones, and an assessment of recovery in the four mixed scenes. All comparisons use the accepted-event budgets and normalization specified above. Full-field images and enlarged regions are displayed on a common intensity scale to compare the fuel-block arrangement, boundary continuity and local texture. High-frequency maps and a common intensity profile further characterize the schedule comparison.

For the backbone study, peak signal-to-noise ratio (PSNR), structural similarity (SSIM) \cite{Wang2004SSIM} and intersection over union (IoU) are calculated against the paired LE reconstruction. The mean and standard deviation of each metric are reported over five SE realizations of the complete scene used for backbone selection. Section~\ref{ssec:ablation} explains how the statistical texture of PoCA images informs the use of visual and numerical comparisons.

All metrics are computed over the full $680\times680$-pixel images. PSNR and SSIM use a data range of 1.0. SSIM is computed with a uniformly weighted $7\times7$ window and stability constants $K_1=0.01$ and $K_2=0.03$. For IoU, the prediction and reference images are binarized using a fixed threshold of 0.4.

\FloatBarrier
\section{Residual diffusion restoration}
\label{sec:method}

\subsection{Observation-conditioned residual states}
\begin{figure}[tbp]
\centering
\includegraphics[width=0.80\textwidth]{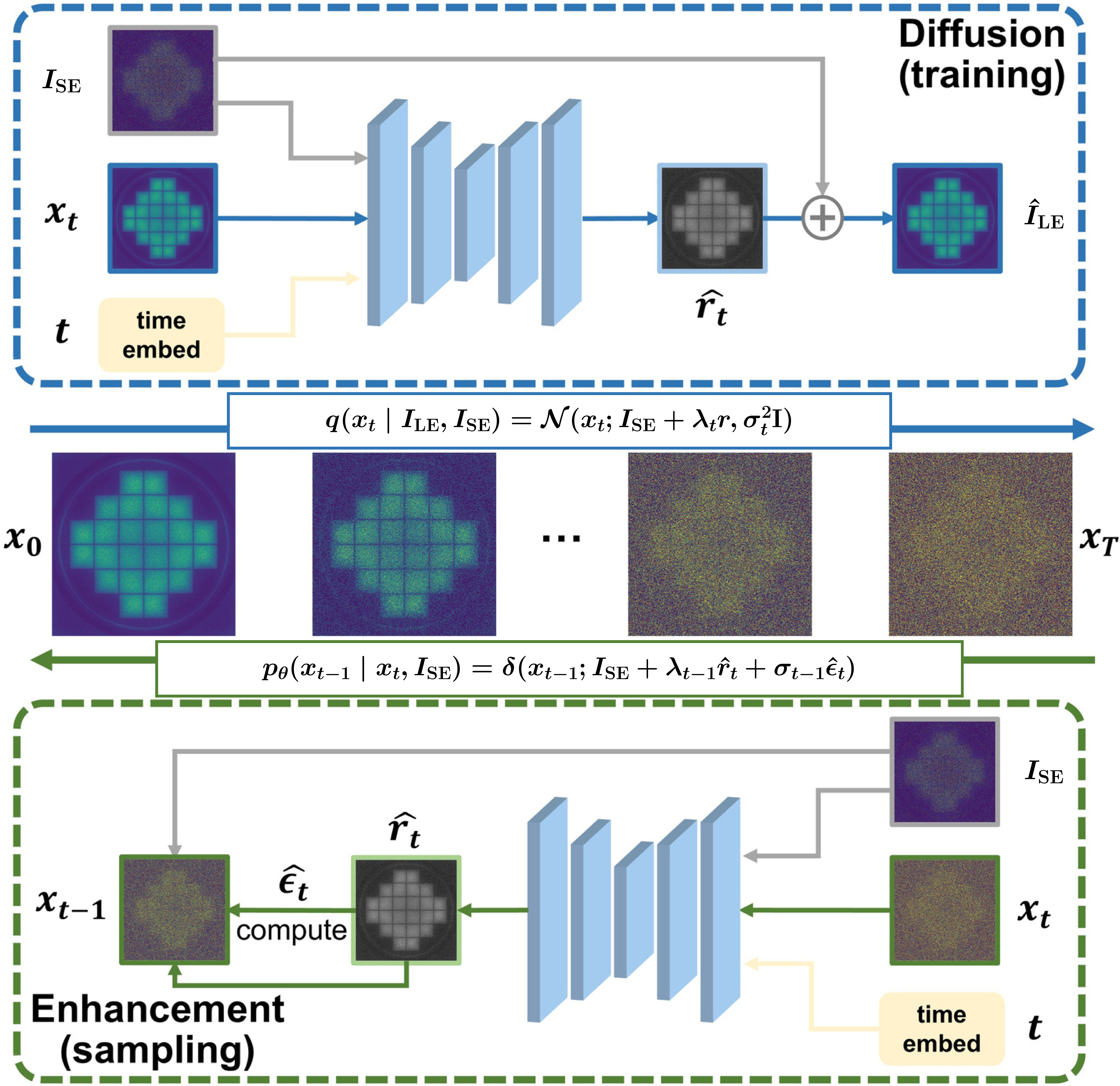}
\caption{Observation-conditioned residual diffusion. Training states interpolate between the normalized images $\ILE$ and $\ISE$ while adding noise; restoration starts from a noisy version of $\ISE$ and iteratively predicts the residual.}
\label{figure3}
\end{figure}

The input image $\ISE$ contains the reconstructed scattering-intensity pattern, while the paired reference image $\ILE$ supplies a statistically richer target. The restoration model learns the residual between the paired images and conditions every sampling step on $\ISE$. Figure~\ref{figure3} summarizes this construction.

For one normalized image pair $(\ISE,\ILE)$ from the same scene, we define the residual as
\begin{equation}
    r = \ILE - \ISE.
\end{equation}

The residual defines a path between the paired images. Gaussian perturbations around this path produce the states used for training:
\begin{equation}
\begin{aligned}
x_0 &= \ILE,\\
x_t &= \ISE + \lambda_t r + \sigma_t \epsilon,\\
x_T &= \ISE + \sigma_T\epsilon,\\
\epsilon &\sim \mathcal{N}(0,\mathbf{I}).
\end{aligned}
\end{equation}

Here, $\mathbf{I}$ is the identity matrix. The endpoint conditions are $\lambda_0=1$, $\sigma_0=0$, and $\lambda_T=0$. The residual coefficient $\lambda_t$ controls the mean state, and the noise coefficient $\sigma_t$ sets the perturbation amplitude. For a fixed image pair, the state distributions are
\begin{align}
q(x_t|\ILE,\ISE) &= \mathcal{N}
\left(\ISE+\lambda_t r,\,\sigma_t^2\mathbf{I}
\right),
\\
q(x_T|\ISE) &= \mathcal{N}(\ISE,\sigma_T^2\mathbf{I}).
\end{align}

The forward mean moves from $\ILE$ toward $\ISE$ as $\lambda_t$ decreases. The terminal distribution remains centred on the observed image $\ISE$. Restoration traverses this path in reverse: the network estimates the residual from the current state, the observed image $\ISE$ and the timestep. The schedules below determine the training states and sampling coefficients.

\begin{figure}[tbp]
\centering
\includegraphics[width=\textwidth]{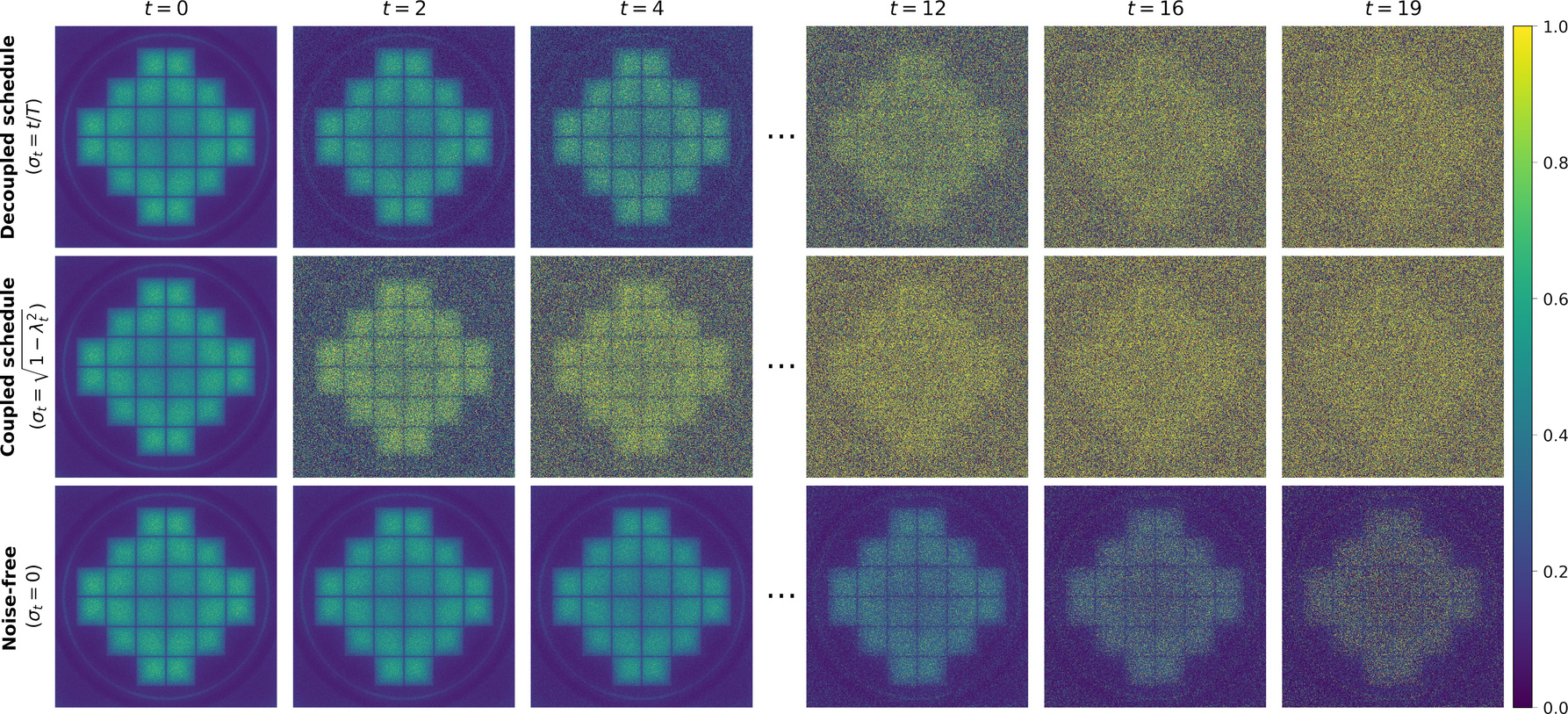}
\caption{Intermediate states along the decoupled, coupled, and noise-free residual trajectories.}
\label{figure4}
\end{figure}

\subsection{Residual and noise schedules}
We use timestep indices $t=0,\ldots,T$ with $T=19$, giving 20 states. The residual is attenuated linearly as
\begin{equation}
\lambda_t = 1-\frac{t}{T}.
\end{equation}
This produces a uniform change in the mean state along the residual path from $\ILE$ to $\ISE$.

To examine how noise growth affects restoration, we first pair the linear residual schedule with a noise coefficient defined by the sum-of-squares relation used in variance-preserving (VP) diffusion:
\begin{equation}
\lambda_t^2+\sigma_t^2=1.
\end{equation}
Applied to the present residual states, this relation defines the VP-inspired coupled variant. The conditional covariance for a fixed LE--SE image pair is $\sigma_t^2\mathbf{I}$. Substituting the linear residual coefficient gives
\begin{equation}
\label{eq:sigma_coupled}
\sigma_t^{\mathrm{coupled}}
=
\sqrt{1-\lambda_t^2}
=
\sqrt{1-\left(1-\frac{t}{T}\right)^2}.
\end{equation}
Letting $\tau=t/T$, for $\tau\ll1$,
\begin{equation}
\sigma_t^{\mathrm{coupled}}
=
\sqrt{2\tau-\tau^2}
\approx
\sqrt{2\tau},
\end{equation}
whereas the residual attenuation is
\begin{equation}
1-\lambda_t=\tau.
\end{equation}
At early timesteps, the coupled noise coefficient therefore grows as $O(\sqrt{\tau})$ and residual attenuation changes as $O(\tau)$. Their contributions to the image are scaled by $\epsilon$ and $r$, respectively.

\begin{figure}[tbp]
\centering
\includegraphics[width=82mm]{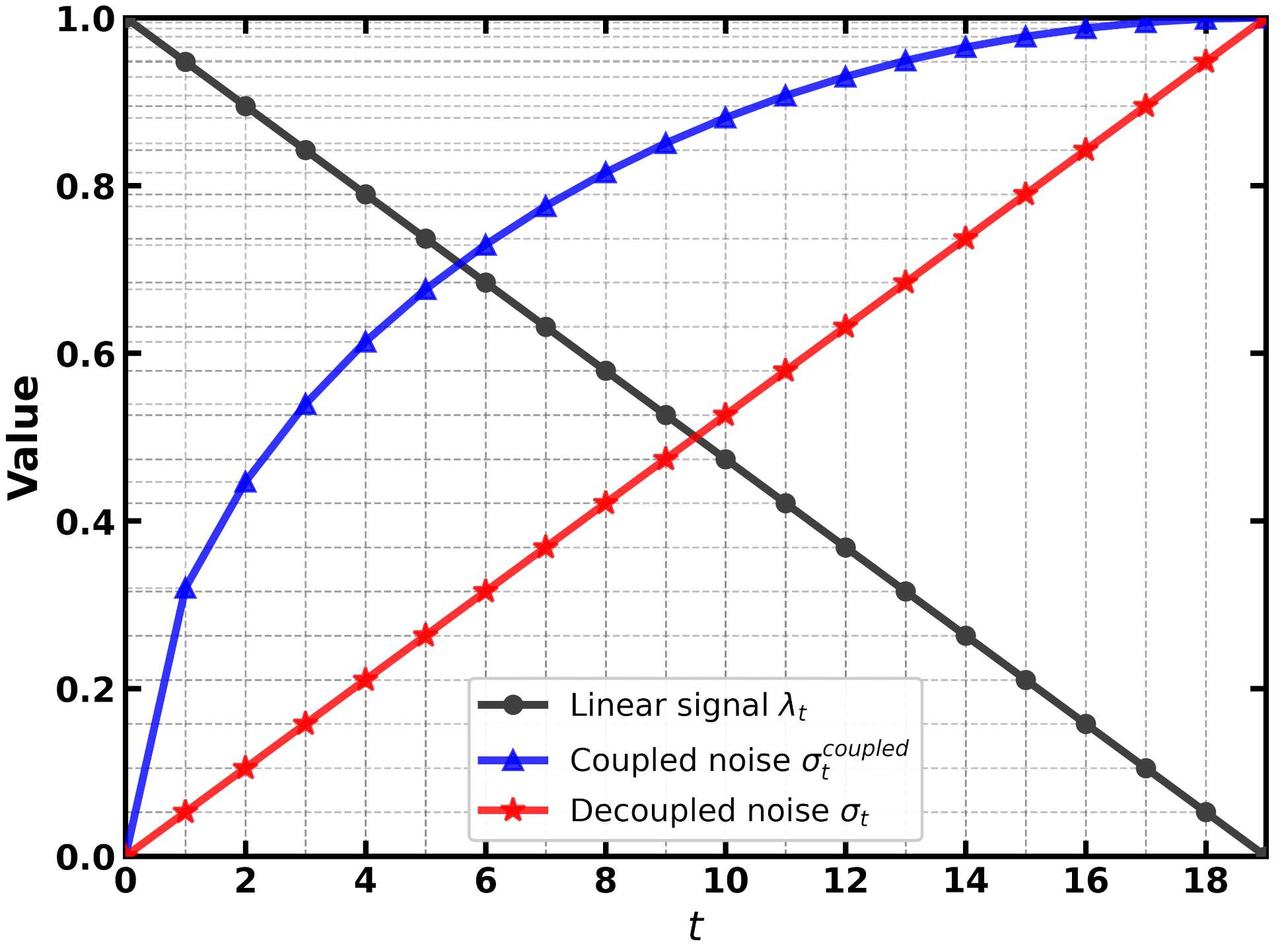}
\caption{Residual and noise coefficients for the linear-decoupled and coupled schedules.}
\label{figure5}
\end{figure}

We remove the sum-of-squares constraint and choose a linear noise schedule:
\begin{equation}
\sigma_t = \frac{t}{T}.
\end{equation}
The proposed forward process becomes
\begin{equation}
x_t = \ISE + \left(1-\frac{t}{T}\right)r
+ \frac{t}{T}\epsilon.
\end{equation}
Residual attenuation and the noise coefficient both vary linearly with $t/T$. The mean trajectory is shared with the coupled variant, while the noise coefficient rises more gradually near the start of the forward process. Here, decoupling removes the sum-of-squares constraint; the selected linear coefficients satisfy $\lambda_t+\sigma_t=1$. This construction permits a direct comparison of how the two noise schedules affect restored boundaries and texture. The effects of this schedule are examined in section~\ref{sec:schedule_comparison}. Figures~\ref{figure4} and \ref{figure5} compare the states and coefficient schedules.

\begin{algorithm}[tbp]
\caption{Training of the residual diffusion model}
\label{alg1}
\begin{algorithmic}[1]
\Require Paired training data $(\ISE,\ILE)$, initialized network $f_{\theta}$, maximum timestep index $T=19$
\Repeat
\State $(\ISE,\ILE)$ from the batch
\State $r=\ILE-\ISE$
\State Sample $t \sim \mathcal{U}\{1,\ldots,T\}$ and $\epsilon\sim\mathcal{N}(0,\mathbf{I})$
\State $\lambda_t=1-\frac{t}{T}$,  $\sigma_t=\frac{t}{T}$
\State $x_t=\ISE+\lambda_t r+\sigma_t\epsilon$
\State $\hat{r}_t=f_{\theta}(x_t,\ISE,t)$, $\hat{I}_{\mathrm{LE},t}=\ISE+\hat{r}_t$
\State Compute $\mathcal{L}_{\mathrm{total}}(\hat{I}_{\mathrm{LE},t},\ILE)$
\State Update $\theta$ by gradient descent on $\mathcal{L}_{\mathrm{total}}$
\Until{converged}
\end{algorithmic}
\end{algorithm}

The stochastic term generates training states around the residual path. Setting $\sigma_t=0$ defines the noise-free comparison:
\begin{equation}
\label{noise-free}
x_t=\ISE+\lambda_t r,
\end{equation}
whose conditional distribution is
\begin{equation}
q(x_t|\ILE,\ISE)
=
\delta
\left(
x_t-\left(\ISE+\lambda_t r\right)
\right).
\end{equation}
Figure~\ref{figure4} shows the resulting deterministic path. Algorithm~\ref{alg2} specifies sampling for the stochastic formulation with $\sigma_t>0$ at each estimated-noise step. For the noise-free variant, sampling starts from $x_T=\ISE$ and follows Algorithm~\ref{alg2} with the noise-estimation step and all noise terms omitted.

\subsection{Training and sampling}
\label{ssec:training}
\begin{table}[tbp]
\centering
\caption{Network and training configurations.}
\label{tab:configuration}
\renewcommand{\arraystretch}{1.12}
\small
\setlength{\tabcolsep}{5pt}
\begin{tabular}{@{}>{\raggedright\arraybackslash}p{0.23\textwidth}>{\raggedright\arraybackslash}p{0.24\textwidth}>{\raggedright\arraybackslash}p{\dimexpr0.53\textwidth-20pt\relax}@{}}
\toprule
\textbf{Parameter} & \textbf{Value} & \textbf{Description} \\
\midrule
input channels & 2 & $[x_t,\ISE]$ as input \\
output channels & 1 & Predicted residual $\hat{r}_t$ \\
channel dims & 32/64/128 & Three hierarchical levels \\
downsample$^{\mathrm{a}}$ & Strided Conv & $4\times4$ Conv with stride 2 and padding 1 \\
upsample$^{\mathrm{a}}$ & Bilinear Conv & Bilinear interpolation + $3\times3$ Conv \\
downsample$^{\mathrm{b}}$ & PixelUnshuffle & Conv + PixelUnshuffle$(2)$ \\
upsample$^{\mathrm{b}}$ & PixelShuffle & Conv + PixelShuffle$(2)$ \\
attention heads$^{\mathrm{b}}$ & 4/4/8 & Three channel levels \\
$T$ & 19 & Maximum timestep index \\
batch size & 2 & Training batch size \\
optimizer & AdamW & Weight decay $1\times10^{-4}$ \\
lr & $1\times10^{-4}$ & Initial learning rate \\
epochs & 200 & Number of training epochs \\
$\lambda_{\mathrm{rec}}$ & 1.0 & Reconstruction loss weight \\
$\lambda_{\mathrm{edge}}$ & 0.5 & Edge-weighted loss weight \\
$\lambda_{\mathrm{range}}$ & 0.01 & Range loss weight \\
\bottomrule
\end{tabular}

\vspace{1mm}
\small{$^{\mathrm{a}}$U-Net; $^{\mathrm{b}}$Restormer.}
\end{table}
\begin{algorithm}[tbp]
\caption{Sampling for SE-to-LE enhancement}
\label{alg2}
\begin{algorithmic}[1]
\Require Observed image $\ISE$, trained network $f_{\theta}$, maximum timestep index $T=19$
\State Sample $\epsilon_T\sim\mathcal{N}(0,\mathbf{I})$
\State Initialize $x_T=\ISE+\epsilon_T$
\For{$t=T,T-1,\ldots,1$}
\State $\lambda_t=1-\frac{t}{T}$, $\sigma_t=\frac{t}{T}$
\State Predict $\hat{r}_t=f_{\theta}(x_t,\ISE,t)$
\State Estimate $\hat{\epsilon}_t=
\frac{x_t-\ISE-\lambda_t\hat{r}_t}{\sigma_t}$
\State $\lambda_{t-1}=1-\frac{t-1}{T}$ and $\sigma_{t-1}=\frac{t-1}{T}$
\State $x_{t-1}
=\ISE+\lambda_{t-1}\hat{r}_t+\sigma_{t-1}\hat{\epsilon}_t$
\EndFor
\State \Return $\predLE=x_0$
\end{algorithmic}
\end{algorithm}

At each training step, we use an image pair, a timestep $t$ sampled uniformly from $1,\ldots,T$, and Gaussian noise to construct
\begin{equation}
x_t = \ISE+\lambda_t r+\sigma_t\epsilon,
\qquad
\epsilon\sim\mathcal{N}(0,\mathbf{I}),
\end{equation}
where
\begin{equation}
\lambda_t=1-\frac{t}{T},
\qquad
\sigma_t=\frac{t}{T},
\qquad
T=19.
\end{equation}
Conditioned on $\ISE$ and $t$, the network estimates
\begin{equation}
\hat{r}_t=f_{\theta}(x_t,\ISE,t),
\end{equation}
and reconstructs the target image through
\begin{equation}
\hat{I}_{\mathrm{LE},t}=\ISE+\hat{r}_t.
\end{equation}
The reconstructed image is compared with $\ILE$ using the loss defined below. Algorithm~\ref{alg1} summarizes the training procedure.

Sampling starts with $x_T=\ISE+\epsilon_T$. At each step, we use the predicted residual to estimate the current noise component, then combine the residual and noise estimates using the coefficients for $t-1$ (Algorithm~\ref{alg2}). This update, analogous to that of a denoising diffusion implicit model (DDIM) \cite{song2022}, is deterministic for a fixed initial noise draw. After 19 updates, $\lambda_0=1$ and $\sigma_0=0$ give
\begin{equation}
x_0=\ISE+\hat{r}_1=\predLE.
\end{equation}
The final output adds the estimated correction to the observed image.

In the training losses below, $\predLE$ denotes the prediction $\hat{I}_{\mathrm{LE},t}$ at the sampled timestep, with the timestep index omitted. The training objective combines an image reconstruction loss, an edge-weighted loss and a penalty for intensities outside the normalized range:
\begin{equation}
\mathcal{L}_{\mathrm{total}}
=
\lambda_{\mathrm{rec}}\mathcal{L}_{\mathrm{rec}}
+
\lambda_{\mathrm{edge}}\mathcal{L}_{\mathrm{edge}}
+
\lambda_{\mathrm{range}}\mathcal{L}_{\mathrm{range}},
\end{equation}
where:
\begin{align}
\mathcal{L}_{\mathrm{rec}}
&=
\frac{1}{N}
\sum_{i=1}^{N}
\left|
\hat{I}_{\mathrm{LE},i}-I_{\mathrm{LE},i}
\right|,
\\
\mathcal{L}_{\mathrm{edge}}
&=
\frac{1}{N}
\sum_{i=1}^{N}
w_i
\left|
\hat{I}_{\mathrm{LE},i}-I_{\mathrm{LE},i}
\right|,
\\
\mathcal{L}_{\mathrm{range}}
&=
\frac{1}{N}
\sum_{i=1}^{N}
\left[
\operatorname{ReLU}\left(\hat{I}_{\mathrm{LE},i}-1\right)^2
+
\operatorname{ReLU}\left(-\hat{I}_{\mathrm{LE},i}\right)^2
\right].
\end{align}
Here, $N$ is the number of pixels, $i$ is the pixel index, and $\operatorname{ReLU}(z)=\max(0,z)$ is the rectified linear unit. The reconstruction loss measures absolute intensity error. Edge weights $w_i$, derived from the normalized Sobel gradient magnitude of $\ILE$, increase the contribution of structural boundaries. The range term penalizes values outside $[0,1]$. Table~\ref{tab:configuration} lists the loss weights and optimization settings.

We implement the residual predictor with either U-Net \cite{ronneberger2015} or Restormer \cite{zamir2022}. Both take the channel-wise concatenation of $x_t$ and $\ISE$ as input and predict a single-channel residual. A sinusoidal timestep embedding, processed by a multilayer perceptron, modulates intermediate features through adaptive scale and shift. The U-Net uses a three-level encoder--decoder with convolutional residual blocks and skip connections. Restormer uses the same hierarchy with multi-Dconv head transposed attention (MDTA) and gated-Dconv feed-forward networks (GDFN). The residual formulation, schedules and training objective are shared, enabling a comparison of the two predictors within a common restoration process.

\FloatBarrier
\section{Results and discussion}
\label{sec:results}

\subsection{Effects of the restoration components}
\label{ssec:ablation}
We compare four restoration configurations using the same U-Net backbone on the complete scene: the proposed decoupled residual diffusion, a coupled-schedule variant, a noise-free residual variant, and direct SE-to-LE mapping. These comparisons examine how the noise schedule, stochastic perturbation and iterative updates affect recovery from the same low-count input.

The four methods differ in both structural detail and statistical texture. A high-count PoCA image contains boundary responses together with point-like fluctuations from a finite sample of scattering events. A restored image can reproduce the spatial arrangement and fluctuation amplitude even when individual texture fluctuations are not aligned pixel by pixel with those in $\ILE$. Such pixelwise differences reduce PSNR, whereas smoothing can reduce the underlying intensity error by averaging the fluctuations. SSIM also responds to local texture statistics, and IoU summarizes mask overlap with limited sensitivity to changes in boundary width or small gaps. These metric sensitivities motivate a visual comparison of the four restoration strategies. We examine boundary continuity, local geometry and characteristic PoCA texture in full-field images and enlarged regions displayed on the same intensity scale. High-frequency amplitudes and intensity profiles provide complementary measures for comparing the schedules.

\subsubsection{Effect of schedule decoupling}
\label{sec:schedule_comparison}
\begin{figure}[tbp]
\centering
\includegraphics[width=\textwidth]{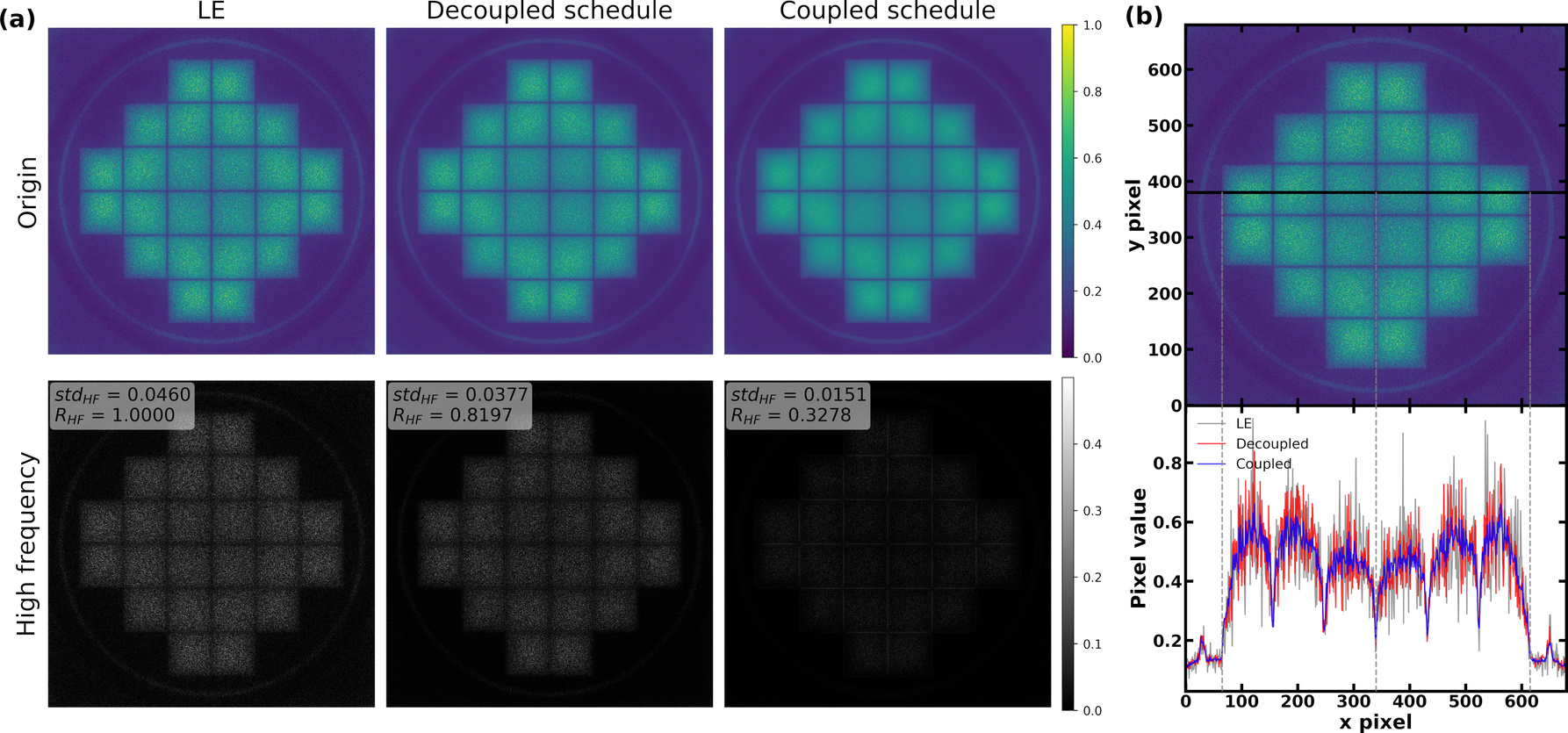}
\caption{Schedule comparison on the complete scene. (a) LE and enhanced images with their high-frequency components. (b) Intensity profiles along the horizontal line marked in the reference image. The high-frequency standard deviation and its ratio to $\ILE$ are shown in each panel.}
\label{figure6}
\end{figure}

The coupled variant retains the linear residual coefficient and uses
\begin{equation}
x_t = \ISE + \lambda_t r + \sigma_t^{\mathrm{coupled}}\epsilon,
\end{equation}
where $\sigma_t^{\mathrm{coupled}}$ is given in eq.~\eqref{eq:sigma_coupled}. Figure~\ref{figure6} compares its output with the decoupled result and $\ILE$. We obtain the high-frequency component of image $I$ by subtracting its Gaussian-smoothed version:
\begin{equation}
I_{\mathrm{HF}}=I-G_{\sigma}(I).
\end{equation}
The same Gaussian filter ($\sigma=2$ pixels, truncated at $4\sigma=8$ pixels with a $17\times17$ kernel and reflected boundaries) is applied to every image; its width is distinct from the diffusion noise coefficient $\sigma_t$.
The high-frequency standard deviation and its ratio to $\ILE$ are
\begin{equation}
\operatorname{std}_{\mathrm{HF}}(I)=\sqrt{\frac{1}{N}\sum_{i=1}^{N}\left(I_{\mathrm{HF},i}-\overline{I}_{\mathrm{HF}}\right)^2},
\end{equation}
\begin{equation}
R_{\mathrm{HF}}(I)=\frac{\operatorname{std}_{\mathrm{HF}}(I)}{\operatorname{std}_{\mathrm{HF}}(\ILE)}.
\end{equation}
Here, $\overline{I}_{\mathrm{HF}}$ is the mean of $I_{\mathrm{HF}}$ over all $N$ pixels. These quantities describe the amplitude of the high-frequency component, including contributions from boundaries and statistical texture.

The high-frequency standard deviation is 0.0460 for $\ILE$, 0.0377 for the decoupled output and 0.0151 for the coupled output, giving ratios to $\ILE$ of 0.8197 and 0.3278 for the two enhanced images. The decoupled schedule retains a fluctuation amplitude closer to the high-count reference. Its high-frequency map shows variations throughout the fuel blocks, whereas the coupled output shows variations concentrated around faint outlines. Along the marked cross-section, both methods recover the sequence of blocks and gaps. The decoupled profile also retains larger local intensity fluctuations, consistent with the closer agreement between the decoupled output and $\ILE$ in within-block texture. Thus, the linear noise schedule improves texture retention while maintaining the broad structural pattern in this comparison.

\subsubsection{Effect of noise perturbation}
\begin{figure}[tbp]
\centering
\includegraphics[width=\textwidth]{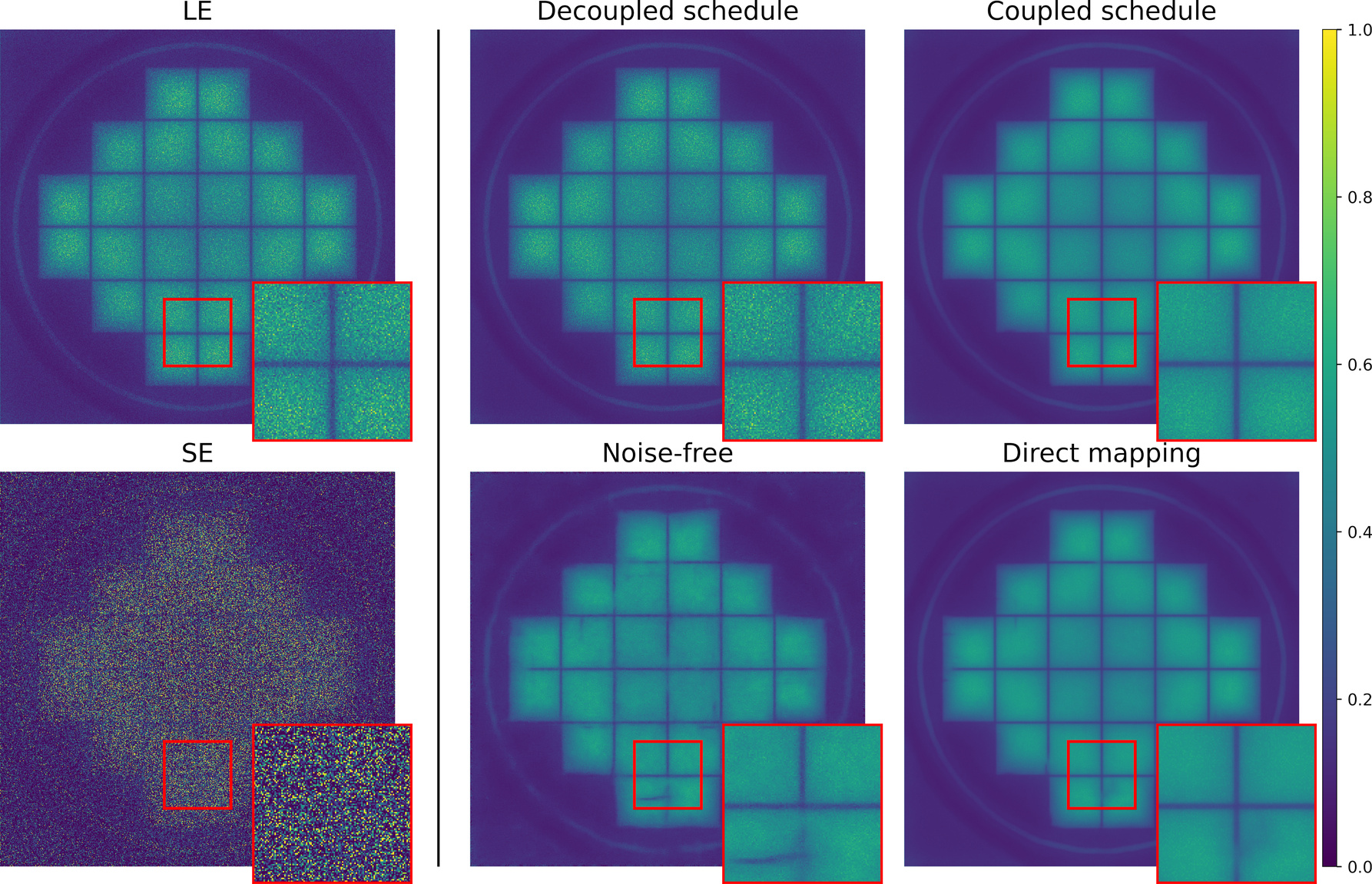}
\caption{Complete-scene PoCA images: LE reference and SE input (left), followed by the decoupled, coupled, noise-free, and direct-mapping outputs. Red boxes indicate the region enlarged in each image. All panels use the displayed common normalized intensity scale.}
\label{figure7}
\end{figure}

Figure~\ref{figure7} shows the four restored images, with the same region enlarged in each. The noise-free variant follows the deterministic residual path in eq.~\eqref{noise-free}. It recovers the main array but produces less regular block outlines, with visible changes in the widths and shapes of the intersecting boundaries in the lower part of the enlarged region. The proposed stochastic formulation produces a more regular arrangement in this region and retains the granular variation within the blocks. The comparison supports the use of perturbed residual states for recovering local structure from the fluctuating SE observation.

\subsubsection{Iterative restoration and direct mapping}
Direct mapping predicts $\ILE$ from $\ISE$ in a single U-Net forward pass, following the strategy used in our previous work \cite{Wang2026}. The residual diffusion model repeatedly updates its estimate using the observation, the current state and the timestep. In figure~\ref{figure7}, direct mapping recovers the overall array with smooth block interiors. The enlarged region shows broader, smoother boundary transitions and changes in the shape of the lower boundary intersection. The decoupled residual model preserves more local variation and produces a clearer boundary intersection. Across the displayed controlled comparisons, the combination of gradual noise growth, stochastic perturbation and iterative correction provides the most consistent recovery of the reference's local boundaries and texture.

\subsection{Comparison of network backbones}
\label{ssec:backbone}
We next compare U-Net and Restormer with the same residual formulation, linear schedules and training objective. The reference-based metrics in table~\ref{tab:backbone_iqa} measure agreement with $\ILE$ under this common restoration setup. Both predictors substantially improve image quality for the complete scene: with U-Net, PSNR increases from 11.4239 to 22.1719 dB, SSIM from 0.0715 to 0.4300, and IoU from 0.2018 to 0.6205. These changes quantify the recovery of an organized fuel-block pattern from the strongly fluctuating SE image.

\begin{table}[tbp]
\centering
\caption{Reference-based metrics on the complete scene. Both enhanced outputs use the residual diffusion framework with linear schedules. Values are mean $\pm$ standard deviation over five SE realizations paired with the same LE image.}
\label{tab:backbone_iqa}
\begin{tabular}{lccc}
\toprule
Image & PSNR (dB)$\uparrow$ & SSIM$\uparrow$ & IoU$\uparrow$\\
\midrule
SE & $11.4239\pm0.0077$ & $0.0715\pm0.0003$ & $0.2018\pm0.0008$\\
Enhanced, U-Net & $22.1719\pm0.0182$ & $0.4300\pm0.0011$ & $0.6205\pm0.0014$\\
Enhanced, Restormer & $22.1860\pm0.0142$ & $0.4294\pm0.0012$ & $0.6215\pm0.0011$\\
\bottomrule
\end{tabular}
\end{table}

The two backbones yield similar mean scores. Restormer yields PSNR and IoU values higher by 0.0141 dB and 0.0010, respectively, while U-Net yields an SSIM value higher by 0.0006. The standard deviations across the five SE realizations paired with the same LE reference are small, indicating stable agreement with $\ILE$ for repeated low-count samples of this scene. The comparable results support U-Net as the predictor used for the subsequent mixed-scene evaluation.

\subsection{Recovery in four mixed scenes}
\label{ssec:mx}
\begin{figure}[tbp]
\centering
\includegraphics[width=\textwidth]{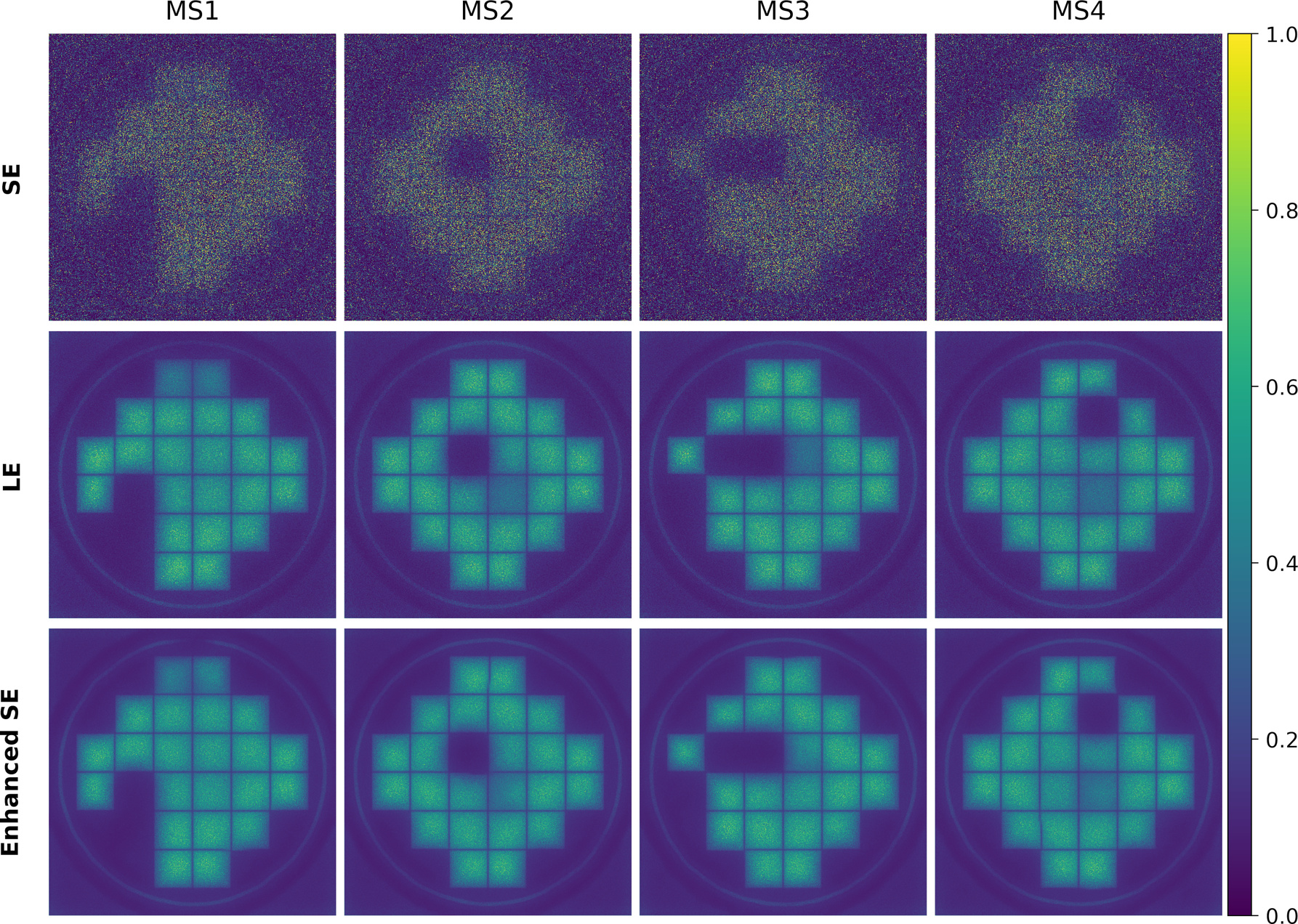}
\caption{Four mixed removal and replacement scenes, MS1--MS4. Rows show SE inputs, LE references and residual-diffusion outputs. All panels use the displayed common normalized intensity scale.}
\label{figure8}
\end{figure}

The four mixed scenes, MS1--MS4, combine removal and iron replacement at different positions within the fuel-block array. They extend the assessment from the complete geometry to configurations with spatially varying material loss and contrast. Figure~\ref{figure8} compares the SE inputs, LE references and outputs of the selected U-Net residual diffusion model.

Strong fluctuations obscure the internal arrangement in each SE image. After enhancement, individual blocks and their intervening boundaries are discernible, and the low-intensity regions follow the scene-specific patterns in $\ILE$. In MS1, the restoration retains the extended low-intensity region on the left side of the lower array. MS2 retains the localized central region; MS3 recovers the broader low-intensity region across the upper-left interior; and MS4 preserves the separated regions in the upper and lower parts of the array. These differences remain visible after enhancement, demonstrating recovery of distinct configurations at the same low event count.

RM regions show pronounced contrast, whereas RP regions exhibit lower contrast in both $\ILE$ and the enhanced images. The restoration recovers the surrounding block arrangement and preserves this contrast pattern, with region contours smoother than those in the LE references. Across the four examples, the method clarifies internal boundaries and altered structures within a metre-scale shielded object, improving the structural information recovered from a fixed sample of accepted accelerator-muon events.

\FloatBarrier
\section{Conclusion}
We have developed an observation-conditioned residual diffusion method for enhancing low-count accelerator-based muon images. The model learns the residual between a low-count input and a high-count PoCA reference and restores the image in 19 steps using decoupled linear residual and noise schedules. Evaluation on a simplified VSC-24 fuel cask demonstrates the method's applicability to a large, shielded object containing closely spaced internal structures. With $5\times10^5$ accepted events under an idealized 5 GeV parallel beam, the U-Net implementation increases the mean PSNR from 11.4239 to 22.1719 dB, SSIM from 0.0715 to 0.4300, and IoU from 0.2018 to 0.6205, as evaluated against the $4\times10^7$-event reference. Controlled comparisons show improved retention of local boundaries and PoCA texture, and four mixed scenes retain distinct removal and replacement patterns after enhancement.

These results demonstrate improved structural recovery at a fixed event budget, supporting the use of learned restoration to reduce the acquisition demands of accelerator-based muon imaging. The image-level restoration framework could also be adapted to cosmic-ray muon scattering images using suitable paired training data. The same observation-conditioned formulation also offers a route to constructing large training datasets from limited accelerator measurements. When applied to measured beam images, the model could generate measurement-conditioned training samples using different initial noise realizations, combining the observed structure with the image characteristics of high-count reconstructions. Such samples could expand the training data available for subsequent learning tasks.

\acknowledgments
This work was supported by the National Natural Science Foundation of China (Grant Nos. 12105327 and 12475106), the Guangdong Basic and Applied Basic Research Foundation (Grant No. 2023B1515120067), and the Research Program of the State Key Laboratory of Heavy Ion Science and Technology, Institute of Modern Physics, Chinese Academy of Sciences (Grant No. HIST2025CS06). This work was also supported by the Fundamental Research Funds for the Central Universities (Grant No. JZ2025HGTG0252). Computational resources were provided by the Dongjiang Yuan Intelligent Computing Center supercomputing facility.

\FloatBarrier
\section*{Conflict of interest}
The authors declare that they have no conflict of interest.
\bibliographystyle{JHEP}
\bibliography{refs}
\end{document}